# Plasma synthesis of single crystal silicon nanoparticles for novel electronic device applications

Ameya Bapat,[1] Curtis Anderson,[1] Christopher R. Perrey, [2] C. Barry Carter, [2] Stephen A. Campbell,[3] and Uwe Kortshagen[1,♣]

[1] Department of Mechanical Engineering, University of Minnesota, Minneapolis, MN 55455

[2] Chemical Engineering and Materials Science, University of Minnesota, Minneapolis, MN 55455

[3] Department of Electrical and Computer Engineering, University of Minnesota, Minneapolis, MN 55455

## Abstract

Single-crystal nanoparticles of silicon, several tens of nm in diameter, may be suitable as building blocks for single-nanoparticle electronic devices. Previous studies of nanoparticles produced in low-pressure plasmas have demonstrated the synthesis nanocrystals of 2-10 nm diameter but larger particles were amorphous or polycrystalline. This work reports the use of a constricted, filamentary capacitively coupled low-pressure plasma to produce single-crystal silicon nanoparticles with diameters between 20-80 nm. Particles are highly oriented with predominant cubic shape. The particle size distribution is rather monodisperse. Electron microscopy studies confirm that the nanoparticles are highly oriented diamond-cubic silicon.

♣corresponding author email: uk@me.umn.edu

## 1. Introduction

Silicon nanocrystals are widely considered a material with great potential for a wide spectrum of applications and novel devices. A variety of novel electronic devices such as single electron transistors [1], vertical transistors [2], and floating gate memory devices [3-6] have been demonstrated. Silicon nanocrystals are also of interest for applications in solid state lighting. While bulk silicon shows basically no photoluminescence due to its indirect band-gap, strong photoluminescence has been demonstrated for silicon nanocrystals even at room temperature [7]. The size-tunable optical and electronic properties of silicon nanoparticles, in combination with an existing silicon technology infrastructure in the semiconductor industry and the element's low toxicity compared to many of the II-IV and III-V semiconductors, make silicon nanocrystals interesting candidates for a wide spectrum of applications, including light emitting devices (LEDs) [8], quantum dot lasers [7], chemical sensors, and molecular electronics [3, 9]. Intense research [7, 8, 10-23] is also focused on developing solid state light sources based on silicon nanocrystal that could in the long term replace energy-inefficient light sources such as incandescent light bulbs (efficiency ~ 7%) or fluorescent lamps (efficiency ~ 25%).

A wide spectrum of synthesis methods is known to produce silicon nanocrystals. Liquid phase processes are widely established for the production of silicon quantum dots. These methods include anodizing silicon wavers in hydrofluoric acid solution [7, 20, 24, 25], synthesis in inverse micelles [26, 27], synthesis in high temperature supercritical solutions [15, 28], the oxidation of metal silicide [12], and the reduction of silicon tetrahalides and other alkylsilicon halides [29]. Unfortunately, none of the liquid phase approaches has so far resulted in the highly desirable high-yield production of silicon nanocrystals. However, solution based processes offer a number of advantages that include their ability to produce particles with a rather narrow size distribution.

Gas-phase processes are attractive for the synthesis of nanoparticles due to high processing rates that can be achieved through direct gas to particle conversion. A variety of gas-phase processes have been proposed to synthesize silicon nanocrystals. These include nanoparticle formation through high temperature thermal reactions (pyrolysis) of silane in furnace flow reactors [4, 10, 30, 31], the decomposition of silane or disilane through laser light irradiation (photolysis) [32] and the laser pyrolysis using high power infrared lasers [14, 17, 33-35]. Unfortunately, most gas phase processes suffer from rapid particle agglomeration due to the fact that particles are usually electrically neutral. For instance, Borsella et al. [16] report a particle size distribution with particle sizes ranging between 1-100 nm for their laser pyrolysis process. Agglomeration of nanoparticles is a severe problem, since it usually annihilates the desired nanoparticle properties.

Nonthermal plasmas have been known for their ability to produce silicon nanoparticles for almost 20 years, however, it was mainly considered from the viewpoint of a contamination problem in semiconductor processing and solar cell production [22, 36-43]. Hence, the potential of using low-pressure plasmas for the high rate synthesis of silicon nanocrystals has so far remained widely underutilized. The pronounced nonequilibrium environment makes nonthermal plasmas attractive for the synthesis of particles for a number of reasons: 1) Nonthermal plasmas offer the same advantage of efficient direct gas to particle conversion as other aerosol processes. 2) Particles immersed in plasmas are usually unipolarly negatively charged [44-47], based on the much higher mobility of electrons in the plasma compared that of ions. This unipolar charge prevents or strongly reduces particle agglomeration [46-48]. 3) Based on their unipolar negative charge particles can be confined in the plasma synthesis reactor through the ambipolar electric fields [39, 49-53]. 4) Nonthermal plasma systems are widely used in the semiconductor industry so that there is a large potential for rapid adoption of nanoparticle synthesis processes based on plasmas.

Several groups have demonstrated the synthesis of crystalline silicon nanoparticles with nonthermal plasmas. Oda *et al.* have developed a "digital plasma process" [54]. In this process crystalline particles are formed in a silane plasma produced in an Ultra High Frequency discharge and ejected from the plasma through injecting periodic pulses of hydrogen gas [9, 55]. In most of their work this group has focused on producing particles of less than 10-15 nm in size. Crystalline silicon and germanium particles were synthesized in a high-density helical resonator plasma by Gorla *et al.* [56]. Free-standing crystalline silicon particles of about 10 nm were also reported in the study of Viera *et al.* [57, 58]. In these studies, nanocrystals were produced in pulsed plasma systems using silane highly diluted in argon. The authors speculate that argon favors the formation of crystalline particles.

In this paper we discuss a plasma process that is capable of producing single crystal, highly oriented (faceted) and virtually defect-free silicon nanoparticles with a highly monodisperse size distribution. The goal of producing these particles is to use them as building blocks for nanoparticle-based electronic devices such as single nanoparticle transistors. In order to be compatible with current lithography capabilities we aim at synthesizing nanoparticles that are several tens of nanometers in size. In previous studies we had shown that low pressure inductive plasmas can be used to synthesize silicon nanoparticles with excellent control over the particle size and the spread of the particle size distribution function [59]. In this study, particles were synthesized in an inductive GEC reference cell with a planar coil in pure silane at a pressure of 10-12 mTorr, an RF power of 200 W, and an overall plasma volume of approximately 1,200 cm$^{-3}$. (For more details we refer the reader to ref. [59].) Particles were found to be highly monodisperse and nonagglomerated and their size could be sensitively controlled via the plasma-on time. However, the silicon particles that were produced in this system were usually amorphous and not suited for device applications due to their high intrinsic defect density [60].

In an effort to synthesize nanoparticles in a plasma that provided a higher gas temperature and more likely yielded crystalline particles, we switched to a different system that operated at significantly higher power density [61]. In this system, particles were synthesized in a flow-through tube reactor using a 4.7 cm inner diameter and a 5.1 cm outer diameter fused silica tube. The plasma was excited by a 5 turn helical coil and an RF power of about 150-200 W was coupled into a plasma volume of about 100 cm$^{-3}$. The process was typically run in 0.72 sccm of silane (5%) in helium (95%) and 3 sccm of argon, leading to a total pressure of 700 mTorr. Extensive Transmission Electron Microscope (TEM) analysis of the particles obtained showed that this process indeed produced single crystal, virtually defect free particles. Particles were generally faceted and highly oriented. Different crystal shapes were observed including cubic particles and particles with a hexagonal appearance. Unfortunately, the particle size distribution found in this study was rather polydisperse and the great variety of different particle morphologies obtained for the same plasma conditions seemed to make this process ill suited for particle synthesis for device applications. For detailed results of these studies the reader is referred to ref. [61].

In this paper we report results of a recent development of a plasma process that yields single crystal silicon nanoparticles with a mainly well defined cubic shape and a rather monodisperse size distribution. The experimental set-up is discussed in section 2, results are presented in section 3. An interpretation of our observations is given in section 4, and the conclusions are summarized in section 5.

## 2. Experimental Set-Up

A schematic of the system used in this study is shown in Fig. 1. Only slight modifications have been made compared to the system used in our previous study [61], however, these modifications lead to a significantly different plasma behavior. The plasma is produced in a quartz tube with 5.1 cm inner diameter and 23 cm length. RF power of about 200 W at 13.56 MHz is applied to a single ring-like RF electrode, which is positioned about 10 cm from the grounded metal flanges at the downstream end of the discharge tube. The discharge is operated using a dilute mixture of 5% silane in 95% helium and argon. A 2.5-sccm-silane:helium (5:95) flow and 3-sccm argon flow yield a discharge chamber pressure of 1500 mTorr. The plasma is typically run for 90 seconds with a 200W RF power input.

The fused silica discharge tube is attached to a high vacuum (HV) chamber. Particles are extracted through a 1-mm orifice. The pressure in the low pressure chamber downstream of the extraction orifice is $10^{-3}$ Torr. The pressure difference leads to the formation of a supersonic gas jet downstream of the orifice. Particles are accelerated in this jet to velocities of up to 250 m/s and deposited on to the TEM grids by inertial impaction. Cu support grids with an amorphous carbon film (01822-F ultra-thin carbon type–A from *Ted Pella Inc.*) were used for the deposition. The base pressure of the systems prior to plasma operation is $10^{-4}$ Torr in the discharge tube and $10^{-7}$ Torr in the HV chamber.

Conventional TEM was performed using a Philips CM30; high-resolution examination was carried out in an FEI Tecnai F30. In both cases the instrument was operated at an accelerating voltage of 300kV

## 3. Results

Applying RF power to the ring-electrode leads to a bright plasma, shown in Figure 2A, between the RF ring electrode and the downstream grounded flange which acts as the ground electrode. The plasma is seemingly diffuse but exhibits a certain structure with a slightly discontinuous luminosity distribution. However, to the human eye the discharge also has an unstable, "flickering" appearance. For this reason, we decided to image the discharge with a high-speed intensified charge coupled device (CCD) camera which allows framing rates of up to 40,000 frames per second (Kodak Ektapro HS motion analyzer 4540mx ). High speed movies taken with this camera reveal that the discharge is not diffuse, as time-averaged images or the human eye suggest, but that it consists of a filament of high luminosity plasma globules (Figure 2B), which rotates in the discharge tube close to the tube wall with a frequency of about 150 Hz. This rotation is not completely periodic, but it is sometimes interrupted by erratic "jumps" of the filament or by changes of the direction of rotation.

In order to understand where particles are formed in this discharge laser scattering experiments were performed. A He-Ne laser beam was directed parallel to the discharge axis through a window at the upstream side of the discharge tube. The laser beam was scanned in radial direction in steps of 1 mm. The scattering light was detected at ~90° scattering angle with a digital camera (Canon Powershot A70) that viewed the discharge tube side on. The emission of the plasma was largely suppressed by using an interference bandpass filter (Edmund Optics NT43-133, peak: 632 nm, FWHM: 10 nm, peak transmission > 45%). The exposure time for each image was 2 seconds. Individual images were combined in Photoshop to give a compound image of the particle scattering signal. Figure 3 shows the time-average optical emission of the discharge together with the obtained laser scattering signal. Upstream of the region of the constricted plasma filament a diffuse plasma region is visible which is separated from the position of the ring electrode by a darker sheath region. It is obvious that the strongest scattering signal is observed in the region between this dark sheath and the RF ring electrode. The scattering is also concentrated close to the walls upstream of the ring electrode, while no scattering is observed from the discharge center. This result suggests that there is a region of strong particle formation or of particle trapping even upstream of the rotating filament plasma region.

Transmission Electron Microscope (TEM) analysis was performed on the nanoparticles that were collected in the high vacuum. Figure 4 shows typical particles extracted from the plasma. Figure 4A demonstrates that the particles are mainly nonagglomerated with a rather monodisperse size distribution. It can be seem that most of the particles are faceted with the cubic crystal shape being predominant. The faces of the cubes are along [100] planes. The difference in the apparent contrast of the particles is due to small differences in the alignment of the particles with respect to the electron beam. Aligned particles show strong Bragg diffraction and appear dark [62]. The particles are single-crystal diamond-cubic silicon as verified by selected-area electron diffraction. Figure 4B shows a higher resolution image of a typical nanoparticle. The particle has an almost perfect cubic shape. No evidence of dislocations or other planar defects was found. A 1-2nm thin amorphous layer surrounds the particle.

Fig. 5 shows the particle-size distribution as determined from image analysis of the TEM images. Using the software NIH Image [63], the particle diameter is obtained from the TEM images by assuming the particle area to be the projection of a spherical particle. The particle size distribution shown in Fig. 5 is based on the analysis of more than 700 particles. It is well approximated by a Gaussian distribution with a peak diameter of 35.4 nm and a standard deviation of 4.9 nm.

In order to verify that the filamentary constricted mode is indeed responsible for forming the cubic single crystal particles, we have also performed experiments in which we avoided the formation of the rotating filament. This can be achieved by changes of the discharge matching and the distance of the ring electrode to the grounded flanges. Unfortunately, this test is slightly ambiguous since the power coupled into the discharge changes as well. Figure 6 shows typical nanoparticles that are found in this diffuse plasma mode. Particles are generally fractal agglomerates (Fig. 6A). The individual primary particles are of a size of ~10-20 nm and are crystalline, as shown in the high-resolution TEM image in Figure 6B. The distinct cubic shape of the nonagglomerated particles obtained from the filamentary constricted mode was not observed in the diffuse plasma mode.

## 4. Discussion

At this point, neither the formation mechanism for the filamentary discharge nor the particle formation mechanism is completely understood. However, the above results suggest several conclusions.

Concerning the mechanism that leads to the filamentary discharge, we suspect that the filament formation is based on the thermal constriction instability, whose mechanism is described in ref. [64]. Based on this scenario, a local increase of the gas temperature in the discharge leads to a local increase in the reduced electric field strength E/N (with E the applied electric field and N the neutral gas density). The usually exponential dependence of the ionization frequency on E/N leads to a strong local increase of the ionization which causes a rise in the plasma density. This rise in plasma density leads to enhanced Joule power dissipation in the filament and to a positive feedback enhancement of the local gas heating. This scenario is supported by the fact that we observe the filamentary constriction only if a sufficiently high partial pressure of Argon buffer gas is used and the total gas pressure is larger than about 500 mTorr. Both measures, the use of a buffer gas with a low thermal conductivity and the use of high pressure promote local gas temperature enhancements.

With respect to the particle formation mechanism, it is important to point out that Figure 3 suggests that initial particle growth may already occur in the diffuse plasma region upstream of the ring electrode. The region of strong scattering seems to indicate either a region of strong particle growth or of particle trapping. Since the RF ring electrode is powered by RF voltages of several hundred volts, the "dark sheath" observed in Figure 3 could indicate the presence of an RF sheath that is established upstream of the ring electrode through capacitive coupling to the upstream grounded metal flanges. If particles are generated in the upstream diffuse plasma and if they acquire a negative charge as expected for particles larger than a few nanometers [47, 65], it is likely that these particles could be trapped in the time-averaged potential of the RF sheath [50, 51]. It is also important to note that the fact that scattering signal is observed with a low power (~10 mW) He-Ne laser and that the scattering is easily observed with the human eye suggests that the particles producing this scattering signal in this region must be of the size of several tens of nanometers. This observation is significant, since for particles several tens of nanometers in diameter, Brownian diffusion should become unimportant [66]. For instance, for spherical silicon particles 20 nm in diameter, one can estimate the typical diffusion time under our experimental conditions to be about 11 seconds, assuming a typical diffusion length of 1 cm. For larger particles, this diffusion time increases rapidly. In comparison, the transit time of particles through the filamentary plasma region is about 4 seconds for our gas flow rates. Considering that both the scattering signal upstream of the ring electrode and the filament rotation are observed close to the wall, it appears likely that the particles travel through the region of filament rotation. The fact that particles can escape the potential trapping region while the plasma is operated was confirmed by exposing TEM grids to the particle stream only during periods of plasma operation. Particles were observed on the grids, which rules out any suggestion that particles are completely trapped during the plasma operation.

While crystalline nanoparticles were observed both in the constricted filamentary mode and in the uniform mode, the particular properties of the filamentary plasma appear to favor a complete sintering of the particles and the formation of single crystals. The fact that crystalline primary particles are also found in the diffuse plasma is not surprising, since cyrstallinization of amorphous silicon has been observed under the influence of atomic hydrogen even at low temperatures [55, 67]. Simple heat transfer calculations for the diffuse plasma, which also apply to the background gas of the filamentary plasma, show that the gas temperature should not exceed the temperature of the gas entering the reactor by more than about 100 K. This is due to the fact that only a small fraction of the RF energy is expended for heating the neutral gas, as discussed below.

In order to understand the processes in the filamentary plasma, we perform a few simple estimates of the plasma properties within the filament. We estimate the plasma properties in the filament by assuming that each of the plasma globules observed in Fig. 3B is a self-sustained plasma in the sense that an ionization-diffusion particle balance needs to be satisfied for each of the globules. For simplicity, we also assume that the plasma properties are mainly determined by the argon buffer gas. While this assumption may be debatable, it may be justified by the fact that at least part of the silane is already dissociated and converted into particles and film on the reactor walls upstream of the filamentary plasma region. Furthermore, we do not expect the helium carrier gas to play a significant role due to its much higher electronic threshold energies as compared to argon. We furthermore limit our simple estimates to the case of ground state ionization of argon, since a detailed treatment of step-wise excitation and ionization is beyond the scope of this discussion. With these assumptions, the electric field amplitude required for self-sustained globules in the filament can be found from the ionization balance

$$\frac{D_a}{\Lambda^2} = \nu_{ion},$$  (1)

where $D_a$ is the ambipolar diffusion coefficient, $\Lambda$ the diffusion length, and $\nu_{ion}$ is the ionization frequency. The diffusion length can be estimated from Fig. 2B as $\Lambda \approx R/\pi$, with $R$ the globule radius. $D_a$ and $\nu_{ion}$ are obtained from a time-dependent Boltzmann solver using argon cross section data [68, 69]. In addition to the self-consistent electric field that is found by balancing the left and the right hand side of equation (1) (see table 1) the Boltzmann solver enables one to analyze the electron energy balance and to find the importance of the various energy absorption mechanisms. Of particular importance is the total Joule power absorbed per electron

$$\Theta_{tot} = \frac{2e}{3m_e} \overline{\int_0^\infty E(t)^2 \frac{u^{3/2}}{\nu_m} \frac{df_0(u,t)}{du} du},$$  (2)

and the power lost per electron in elastic electron-neutral collisions

$$\Theta_{elast} = \frac{2m_e}{M_a} \overline{\int_0^\infty \nu_m u^{3/2} f_0(u,t) du},$$  (3)

which leads to neutral gas heating, where the bar denotes a time-average over one RF period. Here $f_0$ is the electron energy probability function which is normalized as $\int_0^\infty f_0(u) u^{1/2} du = 1$, $u$ is the electron energy in electronvolts, $E(t)$ is the RF electric field strength [70], $\nu_m$ is the electron momentum transfer collision frequency, $m_e$ the electron mass and $M_a$ the ion mass. As shown in table 1, only on the order of 0.5% of the total absorbed power per electron is expended in electron-neutral elastic collisions, which directly contribute to heating of the neutral gas. Using these quantities, the gas temperature in the globule $T_{gl}$ can be estimated with reference to the background gas temperature $T_0$ from the heat diffusion equation assuming steady state

$$T_{gl} - T_0 = \left( \frac{P_{rf} \Theta_{elast}}{N_{gl} \Theta_{tot} V_{gl}} \right) \frac{R^2}{6k_{Ar}} \,. \tag{4}$$

The term in brackets on the right hand side represents the gas heating rate, with $P_{rf}$ the RF power, $N_{gl}$ the number of globules in the filament and $V_{gl}$ the globule volume. $k_{Ar}$ is the thermal conductivity of argon. In addition, the electron density in each globule can be estimated as

$$n_e = \frac{P_{rf}}{N_{gl} \Theta_{tot}} \,. \tag{5}$$

Results of this simple model are presented in table 1. The plasma densities shown in table 1 are an upper estimate based on the assumption that 100% of the RF power is absorbed by the plasma. In a real experiment the power effectively absorbed may be significantly smaller. It is instructive to note that the gas temperature within the plasma globule is only a few tens of Kelvin above the temperature of the surrounding gas. This estimate appears too low and not consistent with the assumption of a thermal constriction of the discharge. However, our estimate does not include the heating of the gas by interactions with hot particles. In order to estimate the importance of this process information about the density of nanoparticles in the plasma would be required, which is currently not available. The heating of the gas through ion acceleration in the sheaths around the particles and subsequent charge exchange is likely negligible, since the sheath dimension scales with the linearized Debye length [71] which is almost by two orders of magnitude smaller than the mean free path for charge exchange.

Several processes may lead to particle temperatures that exceed the temperature of the surrounding gas [65]. Our simple model suggests that rather high plasma densities may be achieved in the globules. Performing a simple energy balance, one finds that electron-ion recombination at the particle surface may be a substantial heating mechanism for the nanoparticles. Assuming a rotation frequency of the filament of 150 Hz and a globule diameter of 3 mm, a particle is exposed to the rotating filament for ~ 130 µs. The heating time scale for a 40 nm diameter particle is about 30 µs. A simplified steady-state energy balance accounting for recombination heating as well as particle cooling by conduction and radiation reads:

$$n_i v_b A_p E_{ion} \ = \ \frac{1}{4} n_g v_g A_p \cdot \frac{3}{2} k_B (T_p - T_g) + \sigma A_p (T_p^{\ 4} - T_g^{\ 4}) , \qquad (6)$$

with $n_i$ the ion density, $v_b$ the Bohm velocity, $A_p$ the particle surface area, $E_{ion}$ the ionization energy of Argon (15.76 eV), $n_g$ and $v_g$ the density and thermal speed of the gas molecules, $k_b$ the Boltzmann constant, $\sigma$ the Stefan-Boltzmann constant, and $T_p$ and $T_g$ the particle and the surrounding gas temperature, respectively. (The wall temperature can by used instead of $T_g$, however, this radiation term in negligible in either case.) In equation (6) we assume that ions approach the particles with the Bohm flux. Alternatively, one can describe the ion collection through the Orbital Motion Limited theory [72, 73]:

$$n_i v_{th,i} A_p \left( 1 + \frac{|eV_p|}{k_b T_i} \right) \ = \ \frac{1}{4} n_g v_g A_p \cdot \frac{3}{2} k_B (T_p - T_g) + \sigma A_p (T_p^{\ 4} - T_g^{\ 4}) , \qquad (7)$$

with $v_{th,i}$ the ion thermal velocity and $V_p$ the potential of the particle. Results for the particle excess temperatures for both models are given in figure 7. While both approaches are highly debatable and more accurate models may be needed in the future, they show that the particle temperature for the expected plasma densities can be several hundreds of Kelvin higher than the surrounding gas temperature. This elevated particle temperature is likely responsible for the annealing of the particles to form the observed defect free single crystals. Gas-particle interactions may also lead to additional gas heating.

At this point, we can only speculate about the origin for the cubic shape of the nanoparticles. To our knowledge, cubic silicon nanocrystals have so far neither been observed in plasma processes nor in any other gas phase process. However, the cubic shape has recently been predicted as the thermodynamic equilibrium shape for silicon particles with a hydrogen passivated surface [74]. We suspect that the annealing due to particle heating in the constricted filamentary plasma and subsequent slow cooling after particle extraction into the low pressure chamber, allow the particles to find a shape close to their equilibrium shape. However, further studies of this topic are required.

Finally it should be stressed that the cubic shape seems particularly desirable for using silicon nanoparticles in device applications, since it leads to an automatic crystallographic alignment of the nanoparticles with respect to the substrate surface and to large area, planar contact surfaces.

## 5. Conclusion

This work has demonstrated a plasma process that is capable of producing a relatively monodisperse, highly oriented nanoparticles 20-80 nm with a predominantly cubic shape. TEM studies indicate that the nanoparticles are crystalline diamond cubic Si and free of planar or line defects. A thin amorphous layer surrounds the particles. The predominant cubic shape of particles is highly desirable for their use in device applications.

The formation of a plasma constriction leading to a rotating plasma filament seems to be essential for the nanoparticle properties. While highly oriented crystalline particles are found in the filamentary mode, fractal agglomerates of crystalline primary particles are found in the diffuse plasma mode. At present, we interpret this filament as being caused by a thermal constriction instability. Laser scattering studies indicate that particles are already formed upstream of the plasma constriction, but that particles may be annealed in the filamentary plasma region. Simple estimates indicate that the particle temperature can exceed the temperature of the surrounding gas by several hundreds of Kelvin in the filamentary plasma region and that electron-ion recombination at the particle surface is the main particle heating mechanism. This enhanced particle temperature appears to be essential in achieving single crystal particles.

**Acknowledgements**

This work was supported by NSF under NIRT grant DMI-0304211, under grant CTS-9876224 and by the MRSEC Program of the National Science Foundation under Award Number DMR-0212302.

| $R_{gl}(mm)$ | $V_{gl}(mm^3)$ | $E_0(V/m)$ | $\Theta_{tot}(eV/s)$ | $\Theta_{elast}(eV/s)$ | $(T_{gl}-T_0)(K)$ | $n_e(cm^{-3})$ |
|---|---|---|---|---|---|---|
| 1.0 | 4.1 | 5563 | $4.9 \times 10^8$ | $8.2 \times 10^5$ | 47.5 | $5.0 \times 10^{13}$ |
| 1.5 | 14.1 | 3708 | $2.5 \times 10^8$ | $6.5 \times 10^5$ | 47.7 | $2.8 \times 10^{13}$ |
| 2.0 | 33.5 | 2855 | $1.7 \times 10^8$ | $5.6 \times 10^5$ | 47.1 | $1.8 \times 10^{13}$ |
| 3.0 | 113.1 | 2000 | $1.0 \times 10^8$ | $4.6 \times 10^5$ | 43.1 | $9.0 \times 10^{12}$ |

Table 1: Results of the simple model for the gas temperature enhancement and electron density in the plasma globules. The calculations were performed for an RF power of $P_{rf} = 200$ W and 12 globules.

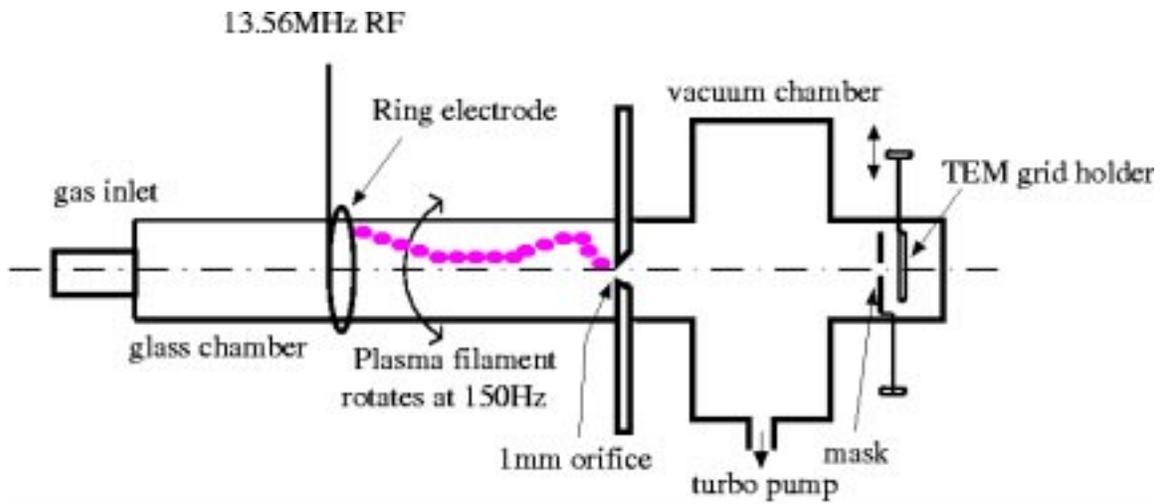

**Figure 1:** Schematic of the experimental set-up.

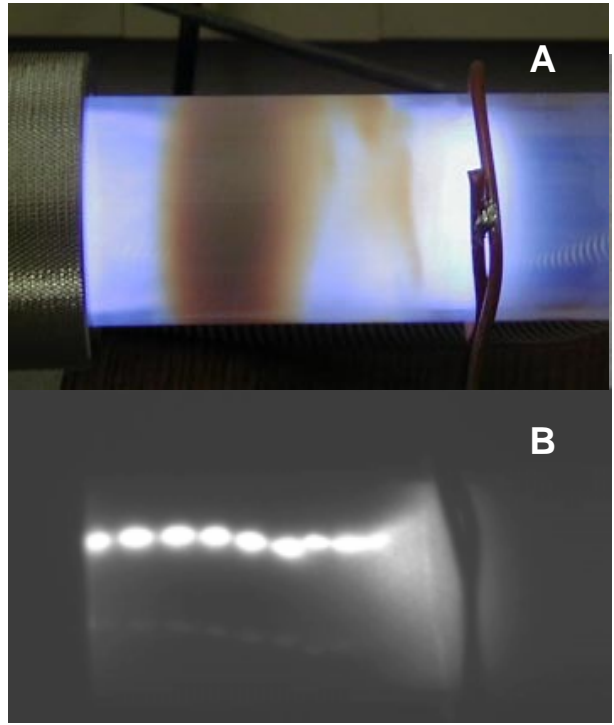

**Figure 2.** Time averaged image (A) and high-speed camera image (B) of the constricted-mode filamentary capacitive discharge.

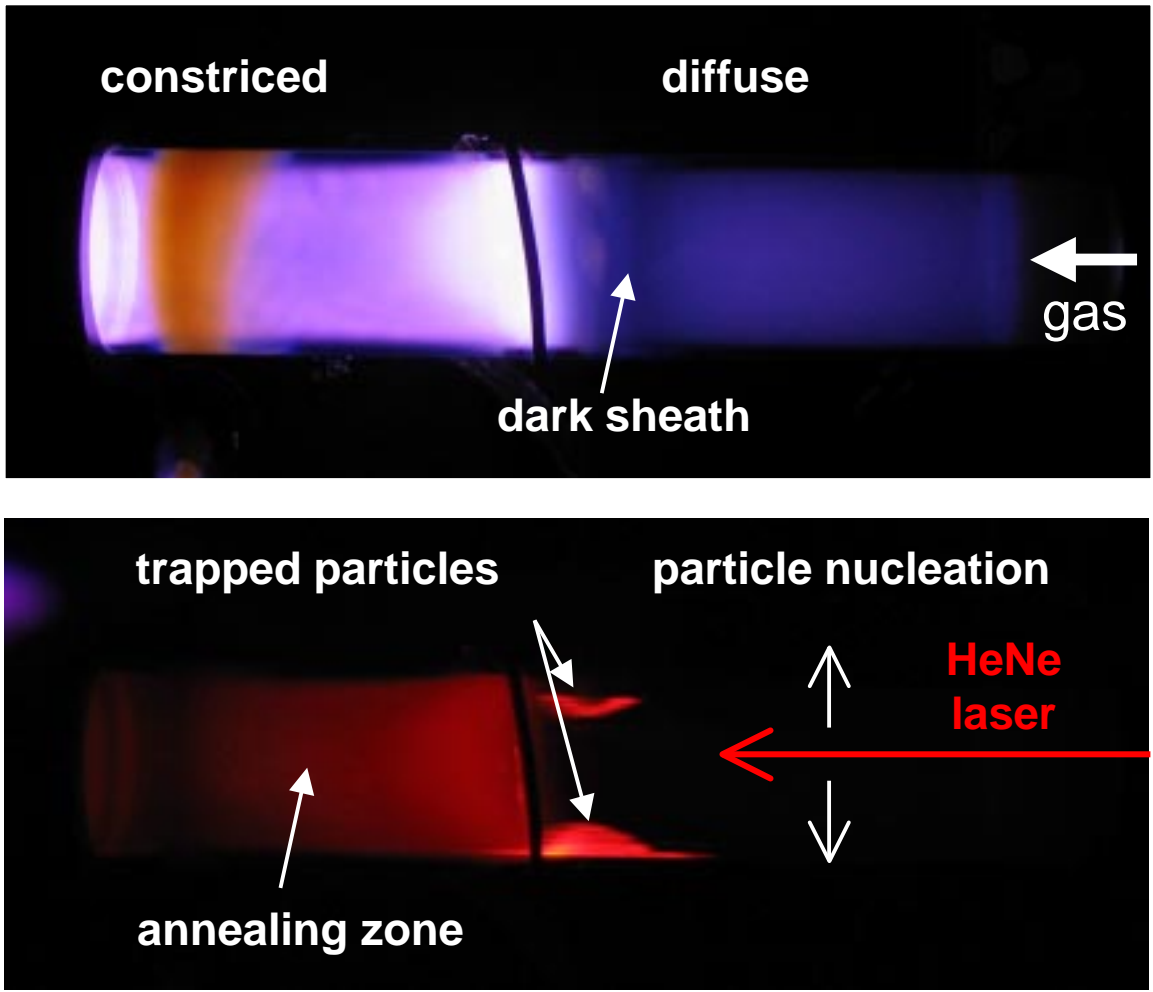

**Figure 3:** Laser scattering signal of particles in the constricted capacitive discharge. The emission observed to the left of the ring electrode is background emission from the plasma and not related to scattering from particles.

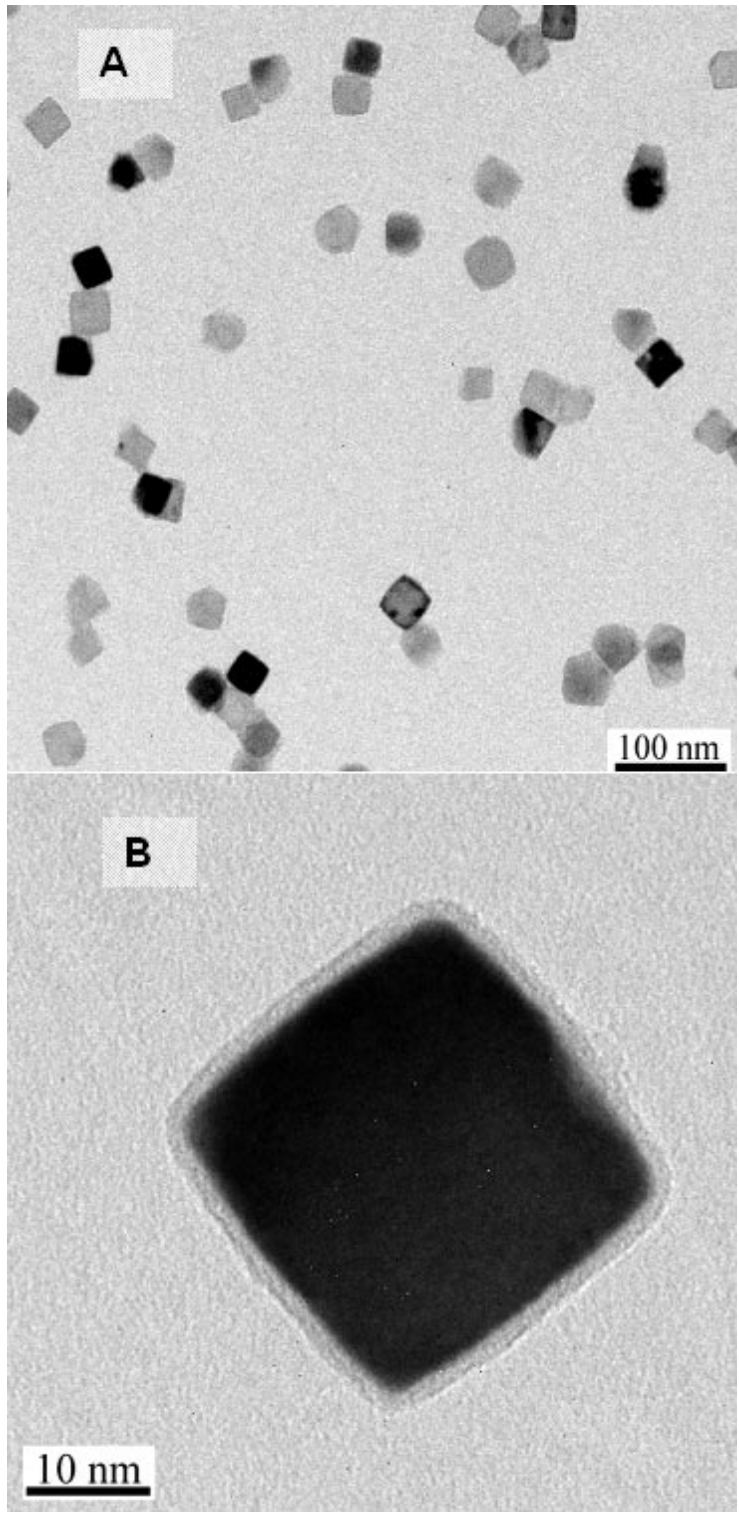

**Figure 4:** Cubic single crystal silicon nanoparticles.

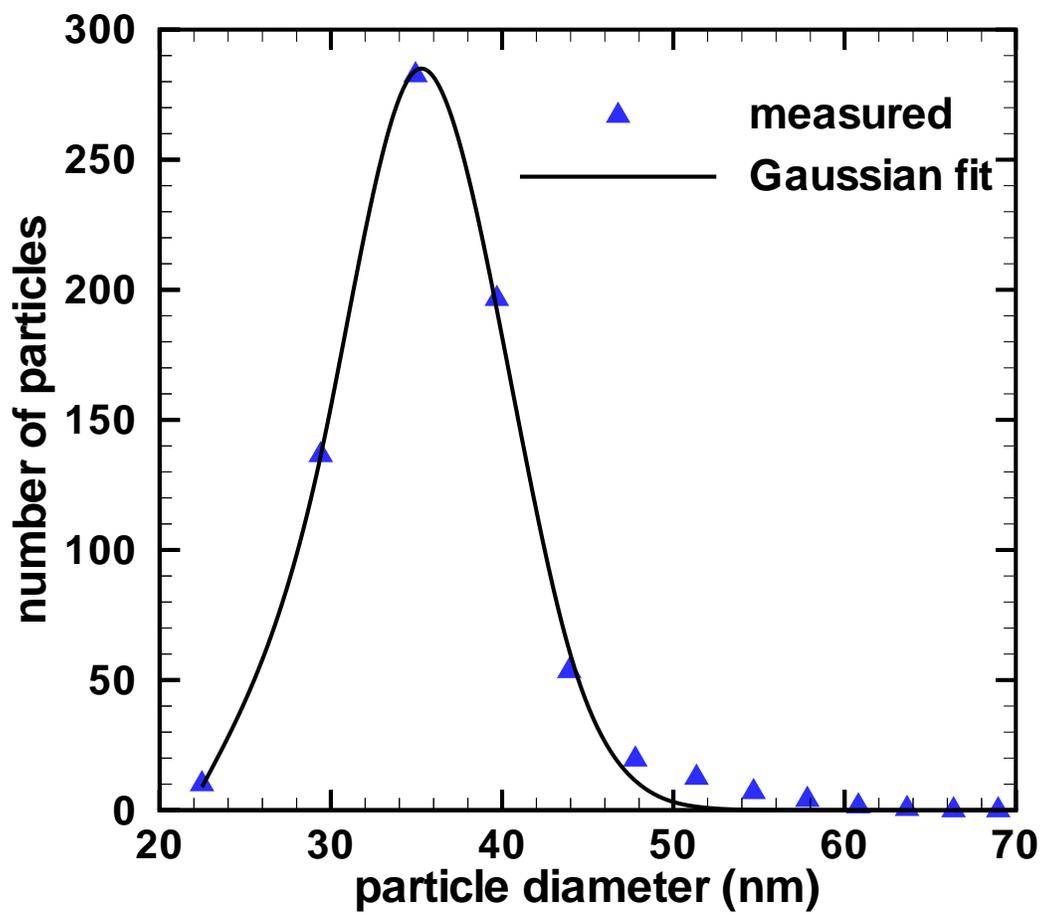

**Figure 5:** Particle size distribution from the constricted mode discharge.

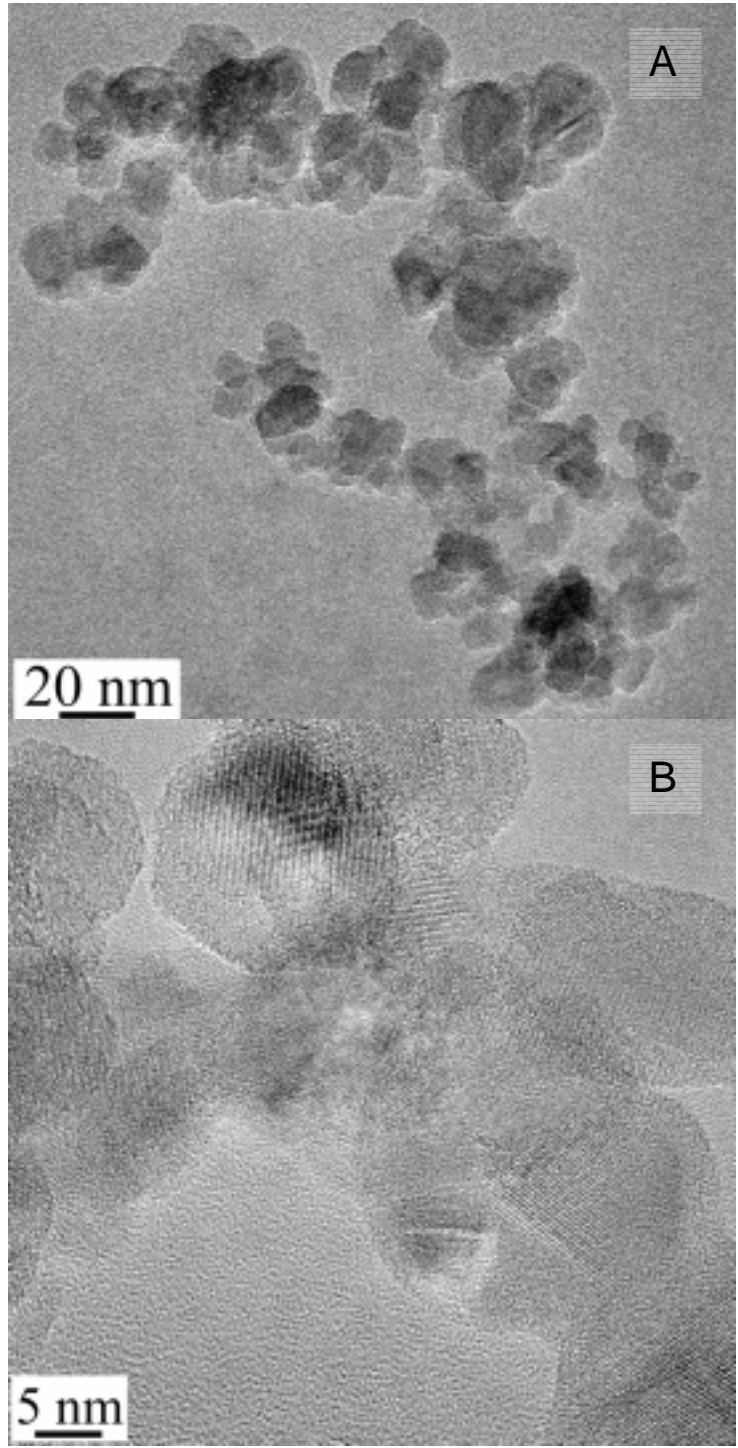

**Figure 6:** Nanoparticle agglomerates formed in the diffuse plasma mode.

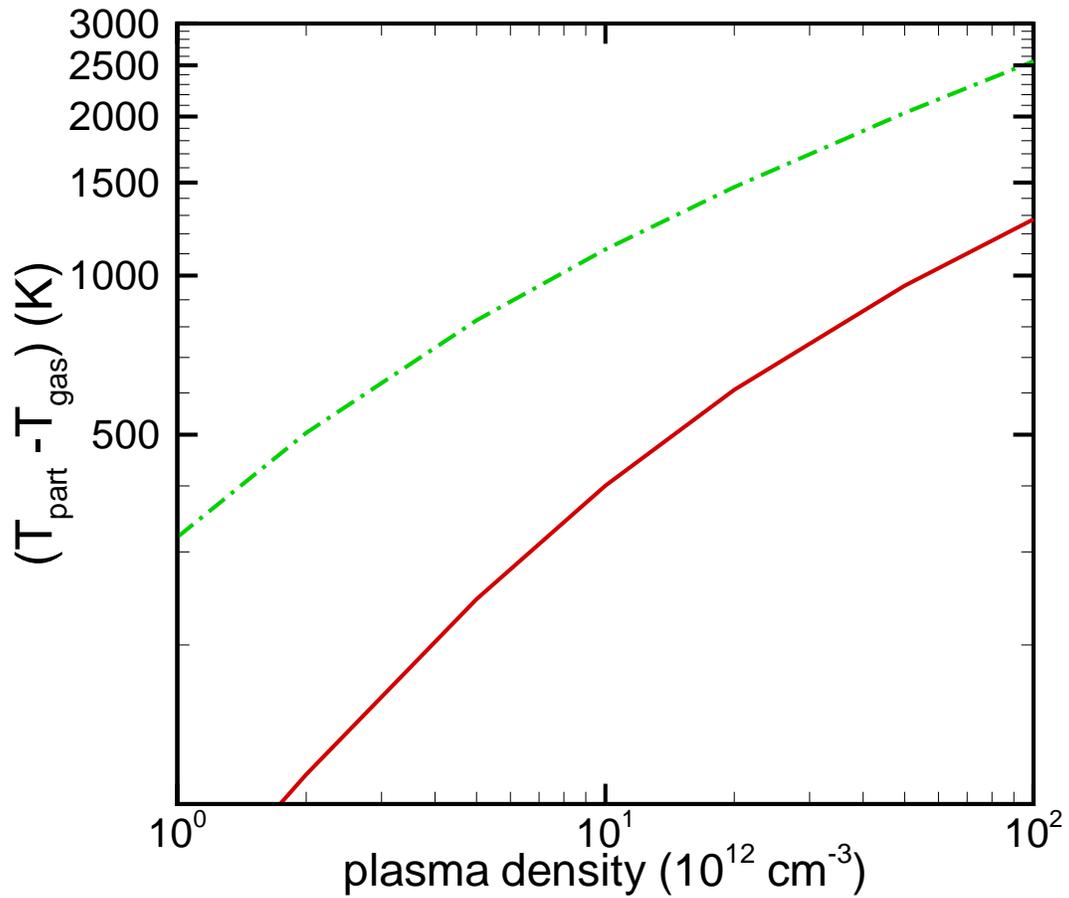

**Figure 7:** Temperature difference between the particle temperature and the neutral gas as a function of the plasma density: The full curve represents ion collection with the Bohm current, the dash-dotted curve represents ion collection following the OML theory.